%% file: main.tex
\documentclass[aps,prb,preprint,superscriptaddress]{revtex4-2}
\usepackage{graphicx}
\usepackage[utf8]{inputenc}
\usepackage[english]{babel}
\usepackage{amssymb,amsmath}
\usepackage{enumerate}
\usepackage{balance}
\usepackage{color}
\usepackage{lineno}
\usepackage{etoolbox}

\makeatletter
\patchcmd{\frontmatter@abstract@produce}
  {\vskip200\p@\@plus1fil
   \penalty-200\relax
   \vskip-200\p@\@plus-1fil}
  {}
  {}
  {}
\makeatother
\begin{document}

\preprint{}

\title{Modelling of time-dependent electrostatic effects and AFM-based surface conductivity characterization}

\author{Mario Navarro-Rodriguez}
 \affiliation{Departamento de F\'isica -- CIOyN, Universidad de Murcia, Murcia 30100, Spain} 
 \affiliation{Experimental Physics of Condensed Matter, University of Potsdam,
Karl-Liebknecht Straße 24-25,D-14476 Potsdam, Germany}
 \author{Paul Philip Schmidt}%
\affiliation{Experimental Physics of Condensed Matter, University of Potsdam,
Karl-Liebknecht Straße 24-25,D-14476 Potsdam, Germany}%
\author{Regina Hoffmann-Vogel}
\affiliation{Experimental Physics of Condensed Matter, University of Potsdam,
Karl-Liebknecht Straße 24-25,D-14476 Potsdam, Germany}%
\author{Andres M. Somoza}
 \affiliation{Departamento de F\'isica -- CIOyN, Universidad de Murcia, Murcia 30100, Spain} 
\author{Elisa Palacios-Lidon}
 \email{elisapl@um.es}
 \affiliation{Departamento de F\'isica -- CIOyN, Universidad de Murcia, Murcia 30100, Spain} 
   
\date{\today}

\begin{abstract}

Atomic Force Microscopy (AFM) combined with electrical modes provides a powerful contactless approach to characterize material electrical properties at the nanoscale. However, conventional electrostatic models often overlook dynamic charge effects, which are particularly relevant for 2D materials deposited on insulating substrates. In this work, we introduce a theoretical framework that extends traditional electrostatic models by incorporating charge dynamics, analyzing two key cases: non-ideal conductors and non-ideal insulators. Our model establishes a characteristic timescale, $\tau$, which governs charge redistribution and measurement reliability. Experimental validation using graphene oxide (GO), reduced graphene oxide, and lightly reduced GO demonstrates strong dependence of frequency shift on surface conductivity, confirming our predictions. Temperature-dependent measurements further reveal conductivity variations consistent with disordered electronic materials. These findings offer valuable insight into how finite surface conductivity influences AFM-based techniques. They introduce a new approach for analyzing charge dynamics on individual flakes of 2D materials while also presenting a contactless method for estimating surface conductivity.
 
\end{abstract}

\maketitle


\section{Introduction}
Understanding electronic processes that take place at nanometer scales is  essential for optimizing nanostructures and thin films in device applications \cite{singh_kpfm_2021,liu_2d_2021,checa_high-speed_2023} as well as in many other research fields such us tribocharging \cite{baytekin_mosaic_2011,pertl_no_2025} and sensing \cite{dufrene_afm_2021,kaswan_recent_2024} where local electrostatic fields and charge redistribution within materials play a crucial role. In addition, charge traps in dielectric materials are an important ingredient to understand device performance \cite{cowie_spatially_2024}. In two-dimensional (2D) materials, their exceptional surface-to-volume ratio and the ability to form homo- and heterostructures through stacking enable the creation of unique and exotic electronic and optical properties, as seen in twisted and Moiré systems \cite{shi_exotic_2021, he_moire_2021}. While crystalline materials like graphene, transition metal dichalcogenides, hexagonal boron nitride, and borophene are prominent examples of 2D materials, disordered materials such as graphene oxide (GO) and reduced graphene oxide (RGO) have also found widespread use in fields like electrochemistry, biomedicine, and sensing \cite{devi_graphene_2024,kiranakumar_h_v_review_2024,waheed_2d_2024,saharan_beyond_2024} thanks to their large surface area, tunable conductivity, and abundant functional groups. A thorough understanding and accurate modeling of the electrical properties of 2D materials, particularly their conductivity, is critical for their incorporation into advanced devices. Fine-tuning of conductivity leads to improved transistor operation \cite{meng_graphene_2025}, more efficient charge separation in photodetectors and solar cells \cite{abid_reduced_2018, dehghan_modification_2021}, and the creation of sensors that rely on conductivity as their sensing mechanism \cite{lee_highly_2019, majhi_reduced_2021}.

Thanks to its high lateral resolution, atomic force microscopy (AFM) and its electrical modes have become essential tools for investigating electrical properties down to the nanometric and atomic scale \cite{meyer_scanning_2021}. Electrostatic force microscopy (EFM) and Kelvin probe microscopy (KPFM) in their multiple variants measure the electrostatic interaction between a biased metallic tip and a sample. This gives access to properties such as contact potential, static or dynamic charge distribution and doping levels, among others \cite{sadewasser_kelvin_2018,glatzel_kelvin_2022}. Their versatility allows for application not only to metallic and semiconductor materials but also to dielectrics and molecular layers \cite{tal_direct_2005,kaja_3d_2024,fumagalli_quantifying_2010,neff_epitaxial_2014,hasan_mechanical_2022,fumagalli_dielectric-constant_2007}. In nanodielectric materials, the ability to measure local dielectric permittivity ($\epsilon_{r}$) and charge accumulation at surfaces or interfaces is particularly important, as these factors play a critical role in device miniaturization and performance \cite{gupta_dielectric_2023-1,cowie_spatially_2024}.

The energy associated to the electrostatic interaction between a biased metallic tip ($V_{\rm bias}$) and a grounded substrate is described by
\begin{equation}
U_{\rm elec}=-\frac{1}{2}C (V_{\rm bias}-V_{\rm SP})^2
\label{elecEnergy}
\end{equation}
where $C(z)$ is the capacitance of the tip-sample system, and $V_{\rm SP}$ is the surface potential. For low and moderate oscillation amplitudes, the corresponding electrostatic force and frequency shift are given by
\begin{equation}
\begin{aligned}
F_{\rm elec}=&\frac{1}{2} C' (V_{\rm bias}-V_{\rm SP})^2,\\
\Delta f_{\rm elec}=&-\frac{f_0}{4 k_{\rm lever}}C''(V_{\rm bias}-V_{\rm SP})^2.
\label{forcefreq}
\end{aligned}
\end{equation}

Both $F_{\rm elec}$ and $\Delta f_{\rm elec}$ exhibit a quadratic dependence on $V_{\rm bias}$ that has been extensively explored through bias spectroscopy. In this technique, these signals are recorded as functions of $V_{\rm bias}$ and the tip-sample distance ($z$) at a fixed sample position. The curvature of the parabolas depends on geometrical parameters as well as on the structure and properties of the sample as they are is determined by the first and second derivatives of the capacitance with $z$. For semi-infinite materials, different capacitance models are employed to fit experimental data depending on whether the material is a conductor or an insulator \cite{sacha_method_2007,sadewasser_kelvin_2018, hudlet_evaluation_1998-1}. These models allow for the determination of relevant parameters such as the tip radius ($R_{\rm tip}$) in metallic samples and the permittivity ($\varepsilon_{r}$) in dielectric materials. They have also been extended to thin films above semi-infinite substrates, broadening their applicability to 2D materials \cite{somoza_localized_2020,pertl_quantifying_2022,labardi_extended_2015}. Typically, a voltage-independent capacitance is assumed, although recent studies have shown that this assumption does not hold in many situations, and in such cases, deviations from a purely parabolic behavior have been used to investigate dangling bonds and band-bending among others \cite{cowie_single-dopant_2022,turek_ring_2020,di_giorgio_imaging_2024}. 

Voltage-independent models for the capacitance do not account for the influence of dynamic effects, in particular charge dynamics at the surface allowed by surface conductivity: when a metallic tip at constant voltage approaches a surface it creates electric fields that originates currents, such that charge reorganizes at the surface in a typical time $\tau$. In absence of bulk conductivity this time is directly proportional to the surface conductivity. In practice, typically only two limits are considered: ({\it i}) metallic limit, $\tau\rightarrow 0$, where it is assumed that charge reorganization occurs on timescales much faster than AFM measurement time, and ({\it ii}) dielectric limit, $\tau\rightarrow \infty$ where any dynamical effect is completely neglected. However, it is expected that if a material exhibits dynamical processes on a timescale comparable to AFM measurements, these effects may significantly influence the tip-sample interaction \cite{checa_high-speed_2023,collins_towards_2018}. In a previous work \cite{navarro-rodriguez_exploring_2024-1}, we have demonstrated that when a 2D material is deposited on an insulating support, surface conductivity can induce topographic artifacts due to surface currents. Moreover, a correlation has been established between the reduction degree of individual RGO flakes and the capacitance-related $2\omega_\textrm{elec}$ signal \cite{navarro-rodriguez_exploring_2024}. This suggests a connection with the conductivity of the 2D material, though its exact contribution remains an open question.

In this work, we investigate the effects of surface charge dynamics on the $\Delta f$ signal for a 2D material on an insulating substrate. We extend the analysis beyond the ideal metallic and dielectric limits by considering both a highly conductive but non-ideal metal and a dielectric with low but non-zero conductivity. In both cases, charge dynamics must be included to accurately describe the capacitance-related term. These effects influence not only the curvature of the parabolic bias dependence but can also lead to deviations from parabolic behavior, especially when the timescale of the charge dynamics matches the measurement acquisition time. This enables a contactless approach to characterize low-conductivity materials where conventional methods are not feasible. We validate our models through AFM measurements on monolayers GO, RGO, and lightly reduced GO, whose conductivity varies by several orders of magnitude and increases with increasing reduction time \cite{pei_reduction_2012-1,eda_chemically_2010}. This tunability ensures charge dynamics occur on timescales relevant to the modeled limits. Furthermore, these materials exhibit hopping conductivity \cite{eda_insulator_2009,joung_efros-shklovskii_2012-2,cheruku_variable_2018-2,park_electrical_2017}. Unlike metals, their conductivity decreases as temperature is lowered due to reduced thermally activated transitions between localized states, allowing for further comparisons with model predictions.

\section{Experimental}

\textit{Sample preparation.} Highly oxidized GO and RGO samples were prepared as described in detail in \cite{navarro-rodriguez_exploring_2024}. Briefly, ultradiluted ($4\times10^{-4}$ wt $\%$) dispersions of GO and and chemically highly reduced RGO in Milli-Q (MQ) water were prepared and drop-cast onto highly doped p-type silicon substrates ($\rho=1-10\ \Omega\cdot$cm) with a $300$ nm thermally grown SiO$_2$ layer. Before deposition, the substrate was rinsed with MQ water and ethanol, then subjected to a $15$-minute UV/ozone treatment to remove organic contaminants and increase the hydrophilicity of the surface. Before transferring to the AFM GO and RGO samples were heated in UHV at $\sim 60^\circ$C for one hour to remove most of the physisorbed water. In addition, one of the GO samples was heated at $100^\circ$C for four hours to achieve light thermal reduction \cite{sengupta_thermal_2018}. From now, the pristine GO and RGO samples are labeled as GO and RGO, respectively while the lightly reduced GO sample is referred to as GO100. 

\textit{Data acquisition.} AFM measurements were conducted under UHV conditions ($p\approx 10^{-10}$ mbar) using a VT-XA SPM (Omicron) equipped with a cryogenic temperature controller (LakeShore Model 336). Metal coated silicon cantilevers (PPP-NCHPt, Nanosensors, $k_{\rm lever}=42$ N/m) with a nominal tip radius $\rm{R_{tip}}=25$ nm were used. The tip was cleaned through multiple cycles of soft Ar-ion sputtering followed by light thermal annealing. Topography images were acquired in non-contact FM-AFM, while KPFM images were recorded in FM-KPFM mode ($\nu_{\rm AC}=330$ Hz, $V_{\rm AC}=500$ mV). The voltage is applied to the tip while the sample is grounded. 

\begin{figure}[h]
    \centering
    \includegraphics[width=8cm]{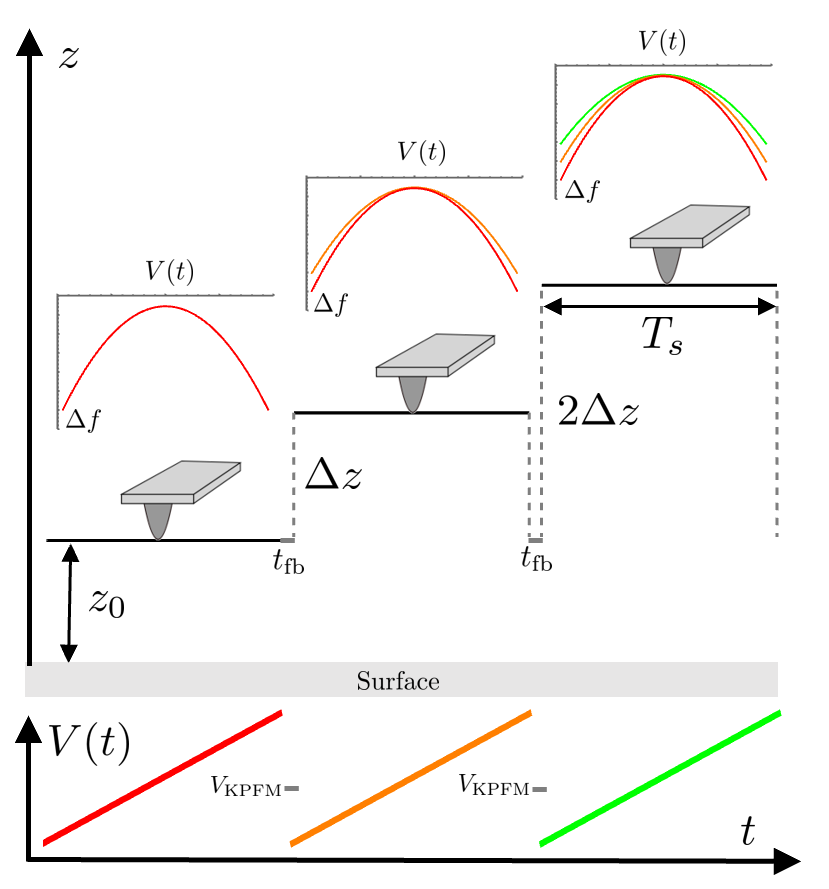}
    \caption{Diagram of the experimental protocol. The horizontal axis represents time. At each tip-sample distance, the frequency shift is recorded as the voltage is swept in a time $T_s$. Between each tip-sample distance the tip returns to the feedback point ($z_0$) with $V_{\rm bias}=V_{\rm KPFM}$ to prevent drift of the tip-sample distance during a time $t_{\rm fb}$ which is represented by the horizontal gray segments.}
    \label{Experimental}
\end{figure}

Frequency shift measurements, $\Delta f$, as a function of the applied bias voltage, $V_{\rm bias}$, were recorded at different tip-sample distances, $z$, at a fixed $x-y$ position. We refer to these $\Delta f(V_{\rm bias}, z)$ images as spectroscopic experiments. The measurement protocol is illustrated in Fig. \ref{Experimental}. The initial tip-sample distance, $z_0$, is determined by the set-point and the tip-sample distance is incremented in steps of size $\Delta z$. At each $z$, the voltage is swept linearly from $V_0$ to $V_{\rm max}$ ($\Delta V \equiv V_0 - V_{\rm max}$) over a time $T_s=5.57$ s in $N_\textrm{points}$ steps, thus $V_{\rm bias}(t) = V_0 + (\Delta V/T_s)t$. The acquisition time for a single point is then $t_a=T_s/N_\textrm{points}$. Before retracting an additional $\Delta z$ step, the voltage is set to $V_{\rm KPFM}$ and the $z$ feedback is reactivated to return the tip to $z_0$. This process ensures minimal drift in the tip-sample distance throughout the experiment and allows the system to return to the electrostatic equilibrium. 

\textit{Data processing.} After acquisition, the $\Delta f ( V_{\rm bias},z)$ curves were fitted for each tip-sample distance using

\begin{equation}
    \Delta f(V,z)= -\frac{f_0(T)}{2k_{\rm lever}(T)}\left(a(z)+C^{\prime\prime}(z)(V_{\rm bias}-V_{\rm KPFM}(z))^2\right).
    \label{freqfit}
\end{equation}

This fitting provides the voltage-independent term $a(z)$, the curvature of the parabola $C^{\prime\prime}(z)$, and the surface potential $V_{\rm KPFM}(z)$ as functions of distance. At each temperature, the resonance frequency $f_0$ and the cantilever force constant $k_{\rm lever}$ were determined from the resonance curve to account for their temperature dependence. The effects of the variation of the geometrical dimensions were neglected \cite{gysin_temperature_2004} and the force constant was determined by the change in resonant frequency \cite{garcia_amplitude_2010} (Section SI in the Supplemental Material \cite{supp}). Aditionally, we confirm that for the  z ranges ($5-60$ nm) and oscillation amplitudes ($\rm{A_{0}}=2-4 nm$) used in our spectroscopy experiments, $\Delta f ( V_{\rm bias},z)$ is independent of the amplitude and the harmonic approximation (Eq. \ref{forcefreq}) remains valid (Section SII in the Supplemental Material \cite{supp}).   

\section{Results and discussion}

In order to study how the conductivity of a 2D material may affect the $\Delta f ( V_{\rm bias},z)$ measurements, we have conducted spectroscopic experiments on individual flakes of RGO, GO and lightly reduced GO (GO100). These three materials cover a wide range of surface conductivities. While GO exhibits a low (but non-zero) conductivity, the light thermal annealing needed to produce GO100 increases its conductivity, though it remains significantly smaller than that of RGO, which is considered a good conductor. Finally, the SiO$_2$ substrate (which is common to all samples) may be considered as an ideal insulator, and serves as a stable reference that is used to control the tip integrity during the experiments.

\begin{figure}[h]
    \centering
    \includegraphics[width=8.5cm]{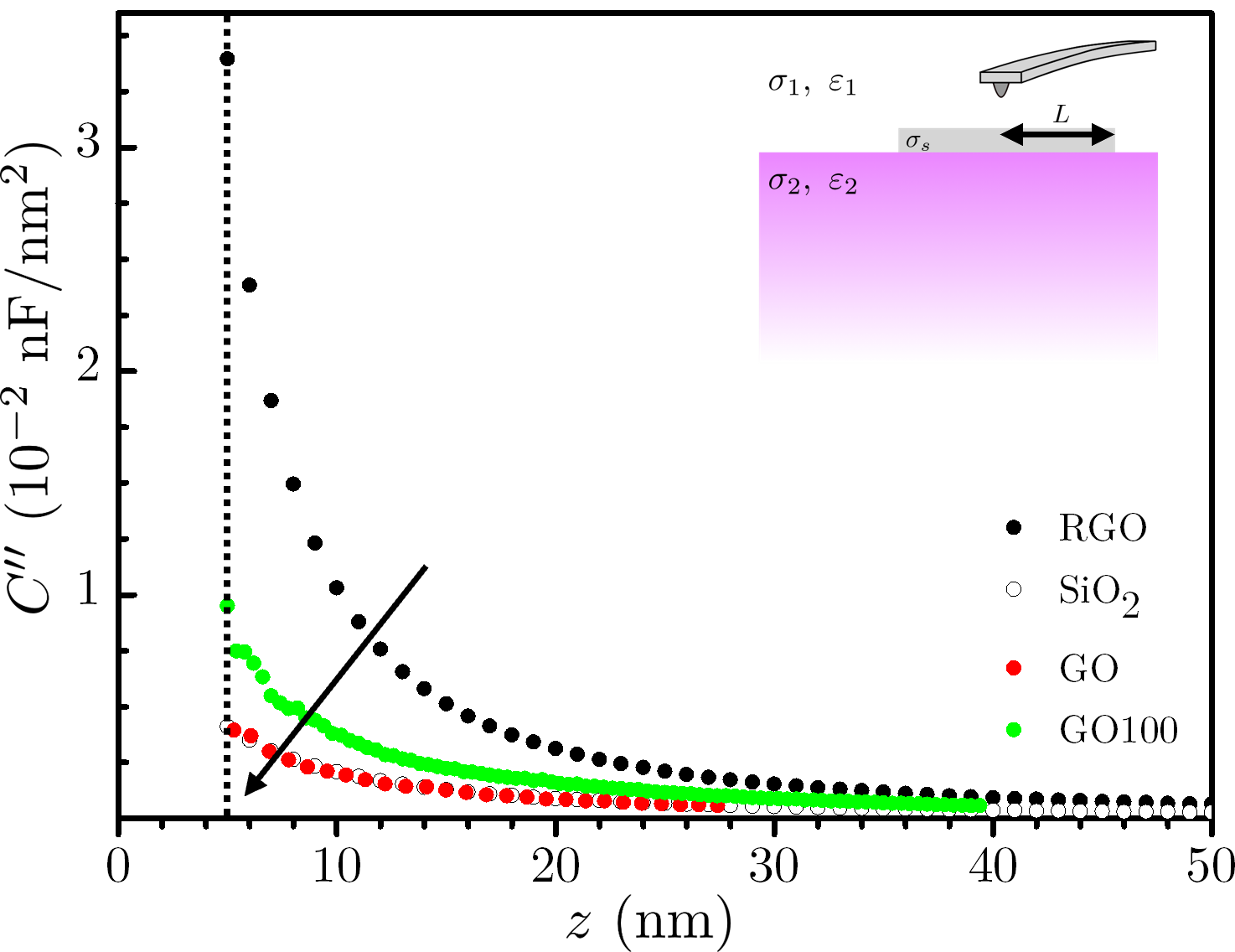}
    \caption{Curvature of the $\Delta f ( V_{\rm bias},z)$ curves for RGO (black), GO100 (green), GO (red), and SiO$_2$ (open dots). All the curves were taken at RT. The inset shows the diagram of the modeled system, a finite size 2D material with size $L$ on top of an insulating substrate. }
    \label{capdiagram}
\end{figure}

As a first approach, we fit the frequency shift $\Delta f(V_\textrm{bias},z)$ to Eq. (\ref{freqfit}) for each of these materials. The corresponding $C''(z)$ term at room temperature (RT) is represented in Fig.\ref{capdiagram}. We observe that there is a clear difference between GO, GO100 and RGO, and it can be argued that there is a monotonic increase of $C''(z)$ with the conductivity. 
The explanation of the results relying on electrostatics alone, assigning an out of plane $\varepsilon_r$ to the 2D material \cite{somoza_localized_2020}, does not lead to an accurate interpretation of our data. If we consider such a model for our data, the required dielectric constant to fit the RGO curve would be of the order of $10^3 \epsilon_0$ (section SIII in the Supplemental Material \cite{supp}). This enormous value for the dielectric constant is out of proportion, even if we consider RGO as pristine graphene. It has been proposed that a graphene sheet can be approximated by a layer of thickness $0.22$ nm with a dielectric constant $\epsilon \approx 6.9 \epsilon_0$ \cite{fang_microscopic_2016-2}, which is much smaller than the value needed to reproduce our data. Even when considering more complex models that account for the dependence of a 2D material’s dielectric response on factors such as the voltage range and probe-sample distance \cite{santos_electric-field_2013-1,ambrosetti_faraday-like_2019-1}, such a large value for the dielectric constant is not predicted. Ultimately, the main conclusion from the literature is that as long as $\varepsilon_r$ is finite, it is reasonable to think that a 2D or ultra-thin material should not have a big influence in the global capacitance of the system. So, in order to explain the behavior found in Fig \ref{capdiagram} new ingredients such as dynamic effects are needed.

In a previous work \cite{navarro-rodriguez_exploring_2024-1}, we derived the equation for the dynamics of surface charges in the presence of an AFM tip at potential $V(t)$ accounting for both bulk and surface conduction. As shown in the inset of Fig.\ref{capdiagram}, the sample is modeled as a semi-infinite medium with permittivity $\varepsilon_2$ and bulk conductivity $\sigma_2$ embedded in a medium with permittivity $\varepsilon_1$ and conductivity $\sigma_1$, where the tip is placed. The interface between the two media is characterized by a surface conductivity, $\sigma_s$. The differential equation that determines the evolution of an arbitrary surface charge distribution is complex. Nevertheless, for our discussion it is enough to consider the limit $z\rightarrow \infty$ as it allows an analytical treatment in Fourier space, giving  

 \begin{equation}
 \begin{aligned}
         \frac{{\rm d}n(k,t)}{{\rm d}t}&=-\frac{n(k,t)}{\tau (k)}+f_{{\rm tip}}(k)V(t),\\
         {\rm with}\  \tau(k)^{-1}&=\tau_0^{-1}+\frac{\sigma_s k}{\varepsilon_1+\varepsilon_2},\\
         {\rm and}\ f_{\rm tip}(k)&=\frac{4\pi R_{\rm tip}}{\varepsilon_1+\varepsilon_2}(\varepsilon_2\sigma_1-\varepsilon_1\sigma_2-\varepsilon_1k\sigma_s)e^{-k z}.
     \label{decay} 
 \end{aligned}
 \end{equation}

Where $n(k,t)$ is the Fourier Transform (FT) of the surface charge density, $\tau(k)$ is the FT of the characteristic decay time with $\tau_0$ the characteristic decay time for bulk conduction. The term $f_{\rm tip}(k)$ depends on the conductivities as well as on the tip geometry, and determines the contribution to the charge dynamics due to the addition of the tip at potential $V(t)$ (see \cite{navarro-rodriguez_exploring_2024-1} for further details). 

Eq. (\ref{decay}) predicts that there are different relaxation times at different lengthscales, with the longest time being the $k=0$ mode, which represents having an infinite surface at constant potential. This characteristic time is then related to the time it takes to establish a constant potential at a surface. For a perfect conductor ($\sigma\rightarrow\infty$) this time tends to zero, while for a perfect dielectric ($\sigma\rightarrow0$) this time tends to infinity. In finite systems, since there is no $k=0$ mode, the $k$ with the largest $\tau(k)$ corresponds to that of the lateral size of the system. For 2D materials on insulating supports, only surface conductivity is considered as $\tau_0\rightarrow\infty$. This is equivalent to setting $\sigma_1,\;\sigma_2\rightarrow0$. Thus, in this limit, the charge decay is entirely mediated by surface conductivity. We note that although Eq.(\ref{decay}) represents a simple particular case, the following discussion remains valid for the exact equations. Below we discuss the metallic and insulating limits capturing the essence of the problem.

\subsection{Metallic limit}

\begin{figure}[ht]
    \centering
    \includegraphics[width=8.5cm]{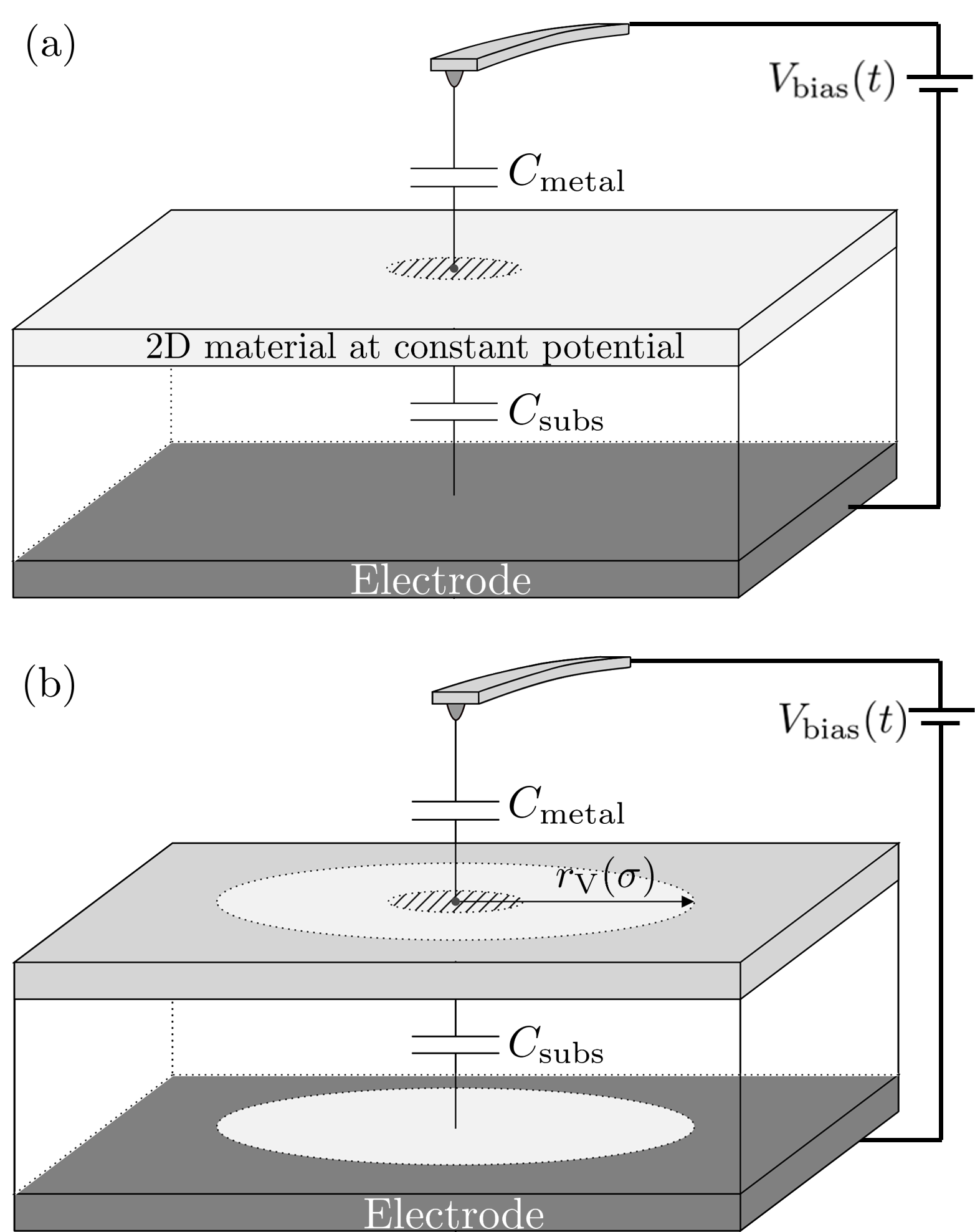}
    \caption{(a) Scheme of the equivalent circuit for an ideal conductor. Light gray represents the area that is at constant potential, which in this limit is the whole flake area and the shaded area represents the area of influence of the tip. (b) For a quasi-ideal conductor, a surface of radius $r_V(\sigma)$ with constant potential, much larger than the area of influence of the tip, acts as the capacitor plate.}
    \label{GoodCond}
\end{figure}

For an ideal 2D conductor, the charge is able to redistribute almost instantly within the conductor to keep the whole flake at constant potential. This constant potential surface can be understood as the plate of a capacitor, and in consequence the total capacitance of the tip-sample system ($C_{\rm tot}$) is given by the sum in series of the tip-flake capacitor ($C_{\rm metal}$) and the flake-back electrode capacitor ($C_{\rm subs}$), as shown in Fig. \ref{GoodCond} (a). In this situation, the total capacitance is written as
\begin{equation}
\begin{aligned}
        C_{\rm tot}(z)&=\frac{C_{\rm metal}(z)C_{\rm subs}(L)}{C_{\rm metal}(z)+C_{\rm subs}(L)}=\alpha(L,z) C_{\rm metal}(z),\\
        {\rm with}\ &\alpha(L,z)=\frac{1}{1+\frac{C_{\rm metal}(z)}{C_{\rm subs}(L)}},
        \label{Cmetlim}
\end{aligned}
\end{equation}
where $L$ is the lateral dimension of the flake. 

For large flakes (typically several microns size), $C_{\rm subs} >> C_{\rm metal}$ as the area of the flake-back electrode capacitor ($\sim \mu\rm{m}^2$) is much larger than the tip-flake capacitor ($\sim {\rm nm}^2$), whose main contribution only comes from a small area around the tip (shaded area in Fig. \ref{GoodCond}). If we reduce the flake area, Eq. (\ref{Cmetlim}) should remain valid as long as the capacitance contribution from the region outside the flake is negligible. However, it is important to note that, as $C_{\rm subs}$ is determined by the area of the 2D material, even if its conductivity is large enough to be considered an ideal conductor, we may find size-dependent capacitance measurements when $C_{\rm subs}(L)\gtrsim C_{\rm tot}$, as $\alpha(L)$ is size dependent. In addition, $C_{\rm subs}$ could also depend on the thickness of the insulating substrate even for metallic flakes.

With this in mind, we can address the quasi-ideal 2D conductor where the surface conductivity is large but not infinite. In this case, one might consider that, during the measuring time the whole surface had no time to reach a constant potential, but shorter relaxation times (corresponding to larger $k$) could certainly have time to adapt. So, as an approximation, we assume that a region of radius $r_V(\sigma)$ below the tip (see Fig.\ref{GoodCond} (b)) was able to reach approximately constant potential, while the rest of the flake remains approximately like in the insulating case. This radius, $r_V$, will become larger if the conductivity is increased. Now, the area of the capacitor $C_{\rm subs}$ is no longer the whole flake, but rather the area with radius $r_V(\sigma)$ at constant potential, thus $C_{\rm subs}=C_{\rm subs}(r_V(\sigma))$, and the total capacitance is

\begin{equation}
\begin{aligned}
    C_{\rm tot}(z,r_V(\sigma))&=\frac{C_{\rm metal}(z)C_{\rm subs}(r_V(\sigma))}{C_{\rm metal}(z)+C_{\rm subs}(r_V(\sigma))}\\
    &=\alpha(r_V(\sigma),z) C_{\rm metal}(z),
    \label{Ctot}
\end{aligned}
\end{equation}

 with $\alpha (r_V(\sigma),z)$ being of order unity since for a quasi-ideal conductor $C_{\rm subs}>>C_{\rm metal}$ still holds. To provide a form of the model that can be validated with experimental data, we calculate the force and frequency shift corresponding to the total capacitance of the system by calculating the first and second derivatives of Eq. (\ref{Ctot})
 
\begin{equation}
\begin{aligned}
    C_{\rm tot}^\prime(z,r_V(\sigma))&=\frac{C_{\rm metal}^\prime(z) C_{\rm subs}(r_V(\sigma))}{C_{\rm metal}(z)+ C_{\rm subs}(r_V(\sigma))}\\
    &-\frac{C_{\rm metal}^\prime(z) C_{\rm metal}(z) C_{\rm subs}(r_V(\sigma))}{(C_{\rm metal}(z)+ C_{\rm subs}(r_V(\sigma)))^2}\\
    &\approx\alpha(r_V(\sigma),z)C_{\rm metal}^\prime(z),
    \label{Cmetaprox0}
    \end{aligned}
\end{equation}

where $\alpha(r_V(\sigma),z)$ is defined as above, and the second term is neglected since, as already justified, $C_{\rm subs}>>C_{\rm metal}$. Similarly, differentiating the previous equation and neglecting the higher order terms in $C_{\rm metal}/C_{\rm subs}$ we obtain

\begin{equation}
\begin{aligned}
    C_{\rm tot}^{\prime\prime}(z,r_V(\sigma))&\approx\frac{C_{\rm metal}^{\prime\prime}(z) C_{\rm subs}(r_V(\sigma))}{C_{\rm metal}(z)+ C_{\rm subs}(r_V(\sigma))}\\
    &=\alpha(r_V(\sigma),z) C^{\prime\prime}_{\rm metal}(z).
    \label{Cmetaprox}
    \end{aligned}
\end{equation}

\begin{figure}[h]
    \centering
    \includegraphics[width=8cm]{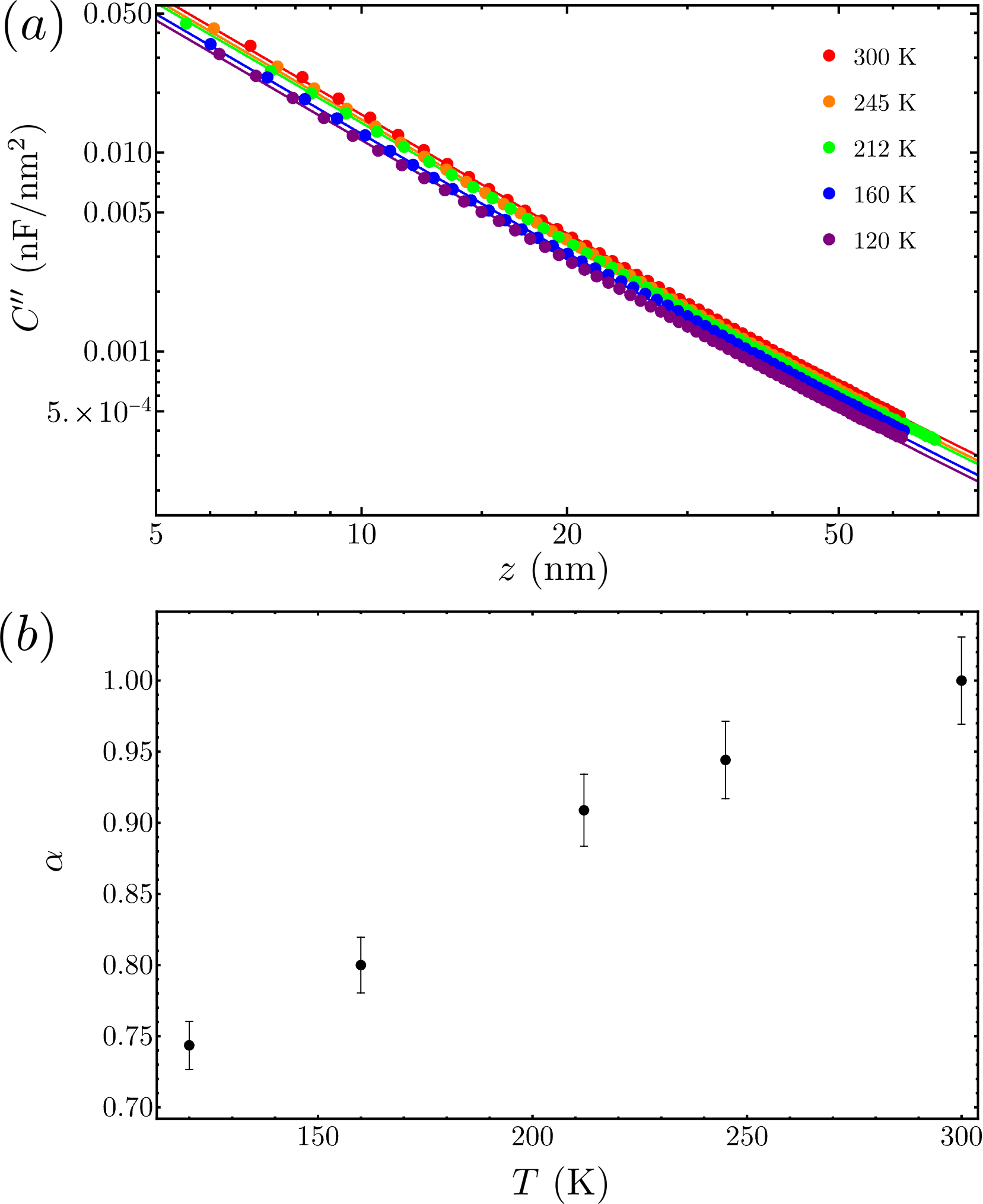}
    \caption{(a) Curvature of the $\Delta f(V_{\rm bias},z)$ curves for rGO as a function of the temperature. The points represent the experimental data, while the solid lines represent the fit to Eq. (\ref{Cmetaprox}). (b) Multiplicative factor, $\alpha$, for every temperature obtained from the corresponding fits.}
    \label{rGOT}
\end{figure}

To explore the good conductor limit experimentally, we have performed spectroscopic experiments on RGO at different temperatures as well as on $\rm{SiO_2}$, that we use as a reference. As mentioned above, since RGO is a disordered electronic material with hopping conductivity, lowering the temperature decreases the amount of thermally activated hops between sites, reducing the conductivity in consequence \cite{eda_insulator_2009,joung_efros-shklovskii_2012-2,cheruku_variable_2018-2,park_electrical_2017}. This temperature-dependent behavior allows to tune the conductivity by cooling the system, enabling to explore the quasi-ideal conductor limit. We can also neglect bulk conductivity as for an air-SiO$_2$ interface at room temperature $\tau_0$ is orders of magnitude larger than our acquisition time \cite{navarro-rodriguez_surface_2023,pertl_no_2025}. Our parabolic fits show that the voltage-independent term does not depend on temperature (Section SIV in the Supplemental Material \cite{supp}). This term is mainly due to Van der Waals interaction, and therefore, we expect the temperature dependence to be negligible. On the contrary, the term $C^{\prime\prime}(z)$ decreases as the temperature is lowered, as shown in Fig. \ref{rGOT} (a). 

Each curve is then fitted to Eq. (\ref{Cmetaprox}) using Hudlet's model for $C_{\rm metal}$ \cite{hudlet_evaluation_1998-1}. This model, widely accepted for the force and frequency shift in metals, predicts that the dependence of $C_{\rm metal}$ with $z$ is logarithmic, and its second derivative is given by

\begin{equation}
\begin{aligned}
    C^{\prime\prime}_{\rm metal}(z)=-2 \pi\varepsilon_0\Bigg(R_\text{tip}\left(-\frac{1}{z^2}+\frac{1}{(z+R_\text{tip}(1-\sin\theta))^2}\right)\\
    -b\left(\frac{1}{z+R_\text{tip}(1-\sin\theta)}+\frac{R_\text{tip}\cos^2\theta/\sin\theta}{(z+R_\text{tip}(1-\sin\theta))^2}\right)\Bigg).
    \label{Hudlet}
\end{aligned}
\end{equation}

Since $C^{\prime\prime}_{\rm metal}$ varies much more rapidly with the distance than $C_{\rm metal}$, we can consider for our comparison that $\alpha$ does not vary with $z$ (Section SV in the Supplemental Material \cite{supp} for further details on the fitting). The corresponding fits (solid lines in Fig. \ref{rGOT} (a)) together with the corresponding values for $\alpha$ obtained at different temperatures (Fig. \ref{rGOT} (b)) show that our model is able to accurately capture the conductivity changes. At RT, $\alpha$ is close to unity, indicating that at this temperature, the behavior of RGO resembles that of an ideal conductor. For the tested temperature range, RGO can still be considered to be close to the metallic limit, and the system is modeled as a metal with a multiplicative constant which encodes the effect of the surface conductivity. In our model, this is interpreted as follows: as the conductivity is decreased by reducing the temperature, the area at constant potential beneath the tip is also reduced, and the ratio $C_{\rm met}/C_{\rm subs}$ increases, leading to a monotonic decrease of $\alpha$. Finally, the fact that we do not observe any temperature dependence on $\rm{SiO}_2$ provides additional confirmation that this trend is due to changes in the conductivity of RGO rather than changes in the tip (Section SVI in the Supplemental Material \cite{supp}).   

\subsection{Dielectric limit}

\begin{figure}[ht]
    \centering    \includegraphics[width=0.45\linewidth]{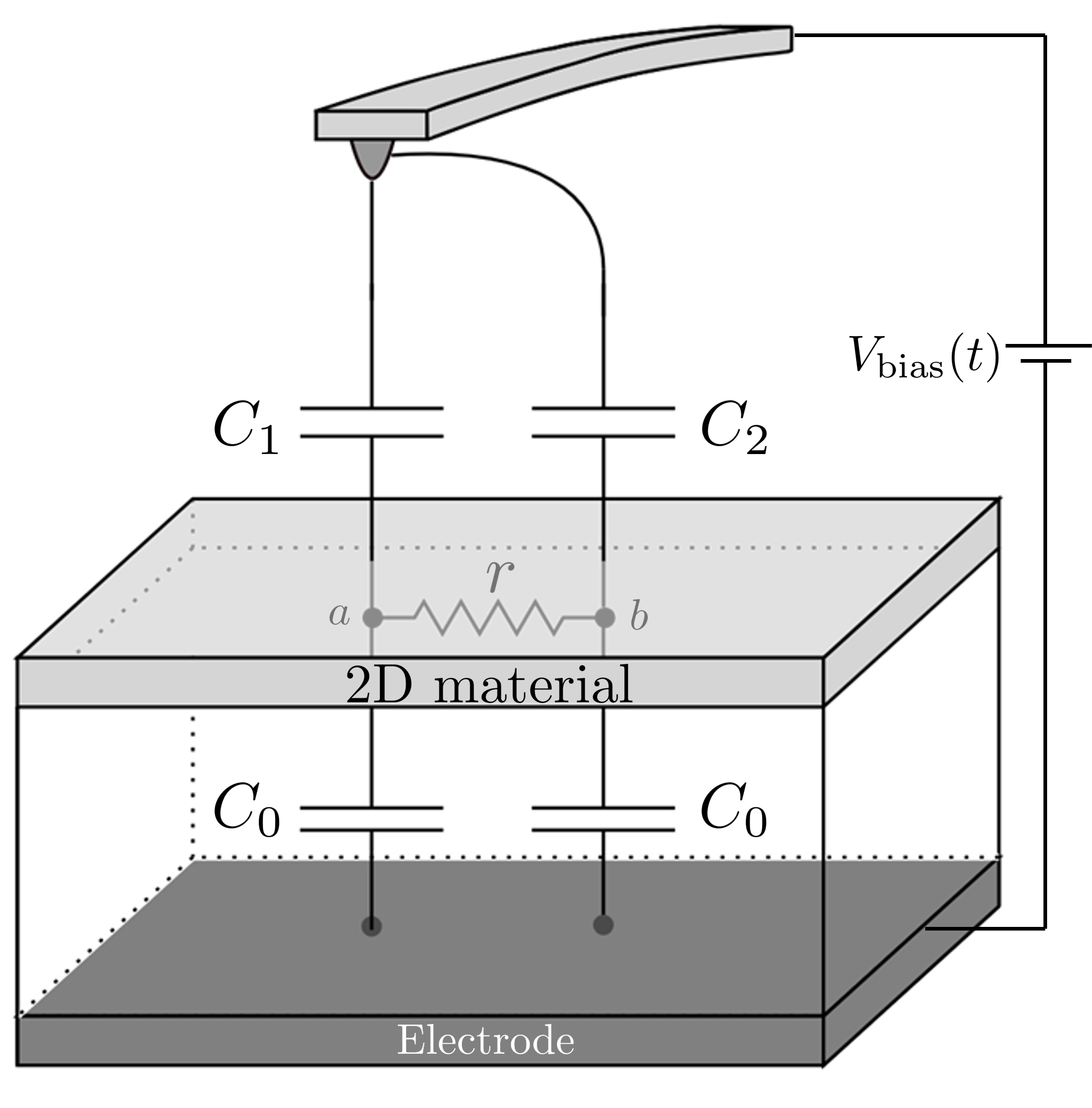}
    \caption{Equivalent circuit for the system of an AFM tip above a 2D material with resistance r on top of a semi-infinite dielectric substrate }
    \label{Circuit}
\end{figure}

 If we now move to the low-conductivity limit, an analysis of Eq. (\ref{decay}) reveals that $f_{\rm tip}(k)$ decays exponentially for $k>1/z$. Hence, for low conductivities, we might assume that dynamics are mainly dominated by a single relaxation time, specifically, the shorter relaxation time with non negligible weight in the dynamics, $k\approx 1/z$. When a bias voltage is applied to the tip, there will always be a component of the electric field parallel to the surface, which induces currents at the surface. Since directly evaluating the electrostatic energy is complex, we instead consider the simplest equivalent circuit, shown in Fig. \ref{Circuit}, which effectively captures the key characteristics of the problem.

In this equivalent circuit, $r$ is the resistor between points a and b and the two capacitors $C_1$, $C_2$ simulate the fact that the potential generated below the tip at two different points in the sample is not the same. Finally, the $C_0$ capacitors represent the capacitance of the substrate below. 

First we consider the ideal dielectric limit where $r\rightarrow\infty$. The charge at each of the elements of the circuit can be written as

\begin{equation}
    \begin{aligned}
   Q_1&=C_1(V_\textrm{bias}(t)-V_a),\; &Q_2&=C_2(V_\textrm{bias}(t)-V_b)\\
    Q_a&=C_0 V_a,\; &Q_b&=C_0 V_b\\
    Q_a&=Q_1,\; &Q_b&=Q_2,\\
    \end{aligned}
    \label{set}
\end{equation}

Where $Q_i$ is the charge stored at capacitor $C_i$, $V_\textrm{bias}(t)$ is the voltage supplied by the battery, and $V_a$, $V_b$ are the potentials at the nodes $a,\; b$ with respect to the ground. In this situation, the total energy of the system is given by 

\begin{equation}
\begin{aligned}
    U&=\frac{1}{2}\left[C_1(V_\textrm{bias}-V_a)^2+C_2(V_\textrm{bias}-V_b)^2\right.\\
    &+\left.C_0(V_a^2+V_b^2)\right]-(Q_1+Q_2)V_\textrm{bias},
    \label{energy}
    \end{aligned}
\end{equation}

where the terms inside the parentheses represent the energy of each capacitor, and the last term accounts for the battery \cite{kantorovich_electrostatic_2000}.

By solving the system of Eqs. (\ref{set}) for $\{Q_1,Q_2,V_a,V_b,Q_a,Q_b\}$ and substituting into Eq. (\ref{energy}), we obtain the energy as a function of the capacitances and the source voltage,

\begin{equation}
    U=-\frac{V^2}{2}\left(\frac{C_0C_1}{C_0+C_1}+\frac{C_0C_2}{C_0+C_2}\right)=-\frac{1}{2}C_{\rm diel}V_\textrm{bias}^2,
\end{equation}

where the terms inside the parentheses correspond to having $C_1$ and $C_2$ each in series with $C_0$. In this limit, each point in the surface is isolated from the rest, and no charge is allowed to flow between different points. Consequently, a constant potential can not be established at the surface. In this limit, the parabolic dependence with the voltage is preserved, and we recover the well-known expression for a tip-back electrode system partially filled with a dielectric layer, being $C_{\rm diel}$ the corresponding capacitance. 

The dielectric limit can be explored experimentally by examining the $\Delta f(V_{\rm bias},z)$ curves for SiO$_2$ and GO. The term $C^{\prime\prime}(z)$, obtained from the fits, yields very similar results for both materials (as was shown in Fig. \ref{capdiagram}), and can be fitted with standard dielectric models \cite{labardi_extended_2015,orihuela_localized_2016} when using the value for $R_\textrm{tip}$ obtained in the metallic limit and a dielectric constant of $\varepsilon_r=4.1\pm 0.3$ (Section SVII in the Supplemental Material \cite{supp}).

\begin{figure}[h]
    \centering
    \includegraphics[width=8cm]{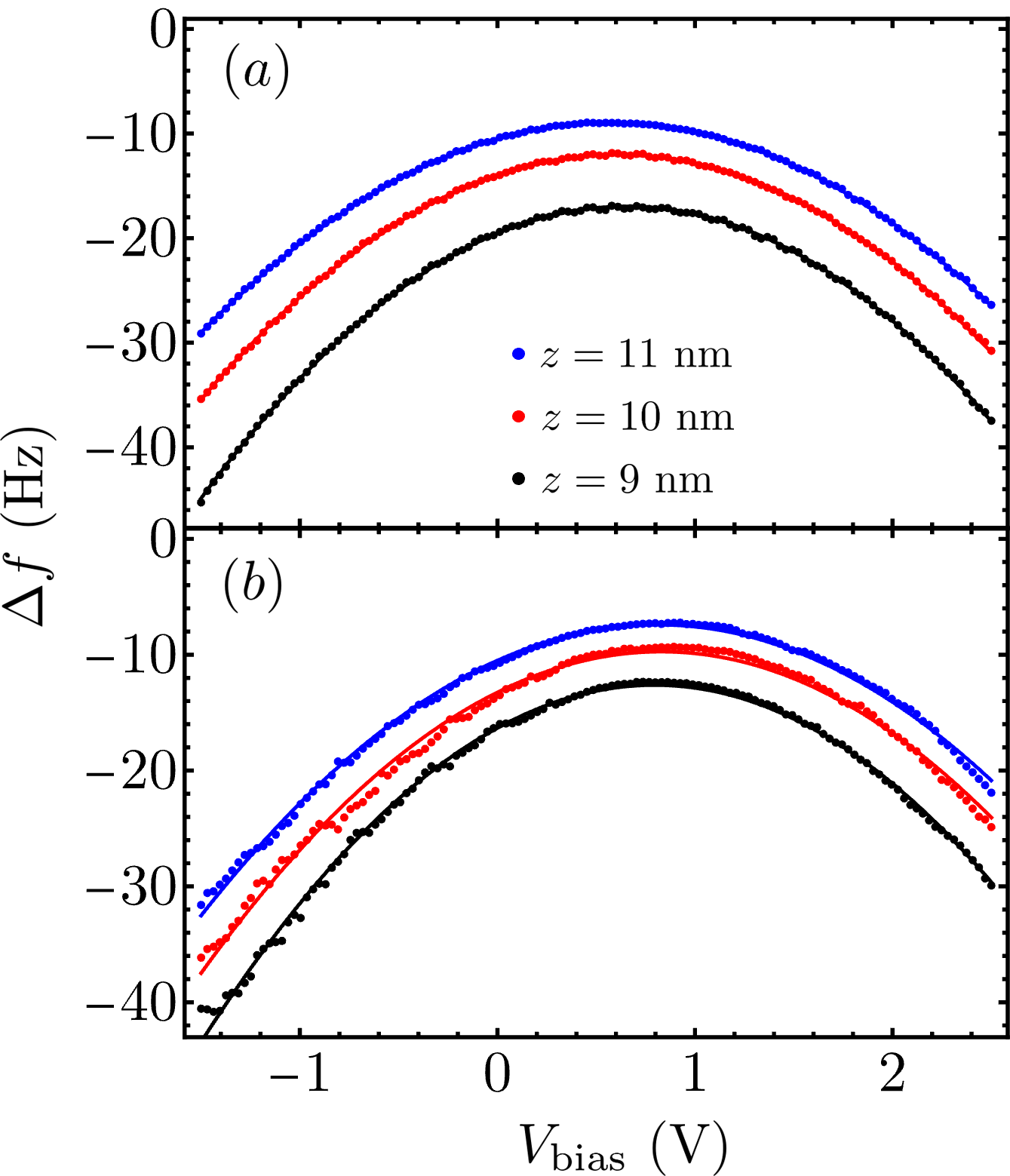}
    \caption{$\Delta f(V_\textrm{bias},z)$ curves for (a) SiO$_2$ and (b) GO, measured at different tip-sample distances. The solid lines represent the fits to Eq. (\ref{freqfit})}
    \label{parabdiel}
\end{figure}

However, a closer examination of the $\Delta f(V_\textrm{bias},z)$ curves reveals a key difference: while the SiO$_2$ curves (Fig. \ref{parabdiel}(a)) closely follow ideal parabolic behavior, the GO curves (Fig. \ref{parabdiel}(b)) exhibit noticeable deviations, particularly at the beginning of the voltage sweep. This discrepancy may arise from GO’s non-ideal insulating nature, where the surface charge distribution is affected by the applied voltage, leading to charge redistribution at the surface.

To include this effect and understand the non-ideal insulator, we assume a large but finite resistance in our model. Now, an amount of charge $\Delta q$ will be exchanged between the nodes $a$ and $b$. As a result, the expressions in Eq. (\ref{set}) must be modified to account for this,
\begin{equation}
    Q_a=Q_1-\Delta q,\qquad Q_b=Q_2+\Delta q,
\end{equation}
Solving the system of equations using the previous expressions for $Q_a$ and $Q_b$, and substituting into Eq. (\ref{energy}) yields

\begin{equation}
\begin{aligned}
   U&= \frac{(2C_0+C_1+C_2)\Delta q^2}{2(C_0+C_1)(C_0+C_2)}\\
   &-\left(\frac{C_0V_\textrm{bias}^2}{2}+\Delta q V_\textrm{bias}\right)\left(\frac{C_1}{C_0+C_1}+\frac{C_2}{C_0+C_2}\right).
   \label{energydq}
   \end{aligned}
\end{equation}

The charge $\Delta q$ will redistribute through the system to reach an equilibrium configuration in a finite amount of time. Once the equilibrium is reached, the value of $\Delta q$ is such that the energy is minimized, ${\rm d}U/{\rm d\Delta q}\rvert_{\Delta q_{\rm equil}} =0$, therefore, 

\begin{equation}
    U=-\frac{C_0(C_1+C_2)}{2C_0+C_1+C_2} V_\textrm{bias}^2.
\end{equation}

Since both nodes have the same potential, the system can be reduced to two capacitors in series, one with capacitance $C_1+C_2$, and the other with capacitance $2C_0$:
\begin{equation}
    C_{\rm tot}=\frac{2 C_0(C_1+C_2)}{2C_0+C_1+C_2}.
    \label{Cmet}
\end{equation}

We note that this capacitance is formally the same as the one obtained in the metallic limit (Eq. (\ref{Cmetlim})) when considering $C_1+C_2\equiv C_{\rm metal}$, and $2C_0\equiv C_{\rm subs}$. However, near the insulating limit one can no longer consider that there is a large region with constant voltage beneath the tip, and the approximation $C_{\rm subs} >> C_{\rm metal}$ used for the capacitance and its derivatives in the conducting limit is no longer valid.

When a time dependent voltage ($V_{\rm bias}(t)$) is applied between tip and sample, charge will flow between the capacitors in order to minimize the energy of the system. However, depending on the conductivity and the rate of change of the voltage, the system might have not reached equilibrium before the voltage changes. The current across the resistor is written as

\begin{equation}
\begin{aligned}
    I&=\frac{{\rm d}\Delta q}{{\rm d}t}=\frac{V_b-V_a}{r}\\
    &=-\frac{(2C_0+C_1+C_2)}{(C_0+C_1)(C_0+C_2)r}\Delta q\\
    &\quad \, +\frac{C_0(C_1-C_2)}{(C_0+C_1)(C_0+C_2)r}V_\textrm{bias}(t).
    \label{diffeq}
    \end{aligned}
\end{equation}

The mathematical structure of this equation is equivalent to Eq. (\ref{decay}), which describes the evolution of an arbitrary surface charge density in the presence of an AFM tip when considering a single dominant decay mode. Assuming $V_\textrm{bias}(t)=V_0+(\Delta V /T_s) t$, as in our experiments, the solution to Eq. (\ref{diffeq}) with the initial condition $\Delta q(0)=0$ is

\begin{equation}
\begin{aligned}
\label{eq:dq}
        \Delta q (t)&=\frac{C_0(C_1-C_2)}{(2C_0+C_1+C_2)^2}e^{-\frac{t}{\tau}}(-V_0(2C_0+C_1+C_2)
        \\&+\frac{\Delta V}{T_s} r(C_0+C_1)(C_0+C_2)) \\   
       &+\frac{C_0(C_1-C_2)}{(2C_0+C_1+C_2)^2}\left[\left(V_0+\frac{\Delta V}{T_s} t\right)(2C_0+C_1+C_2)\right.\\
       &-\left.\frac{\Delta V}{T_s} r (C_0+C_1)(C_0+C_2)\right],
\end{aligned}
\end{equation}

where $\tau^{-1} \equiv (\frac{1}{C_0+C_1}+\frac{1}{C_0+C_2})\frac{1}{r}$ is the characteristic time for charge redistribution at the surface. Substituting the previous expression into Eq. (\ref{energydq}) and neglecting terms that decay faster than $\exp(-t/\tau)$ gives

\begin{equation}
\begin{aligned}
  U(t)=&  -\frac{1}{2}C_{\rm tot}(z) V_\textrm{bias}(t)^2+\mathcal{A}(z)\exp{(-t/\tau)}+\mathcal{B}(z).\\
  {\rm with}\quad  \mathcal{A}(z)\equiv&-\frac{\Delta V}{T_s}\frac{rC_0^2(C_1-C_2)^2}{(2C_0+C_1+C_2)^2}  \left(-V_0 +\frac{\Delta V}{T_s}\tau\right),\\
  {\rm and}\quad \mathcal{B}(z)\equiv&\left(\frac{\Delta V}{T_s}\right)^2\frac{\tau C_0^2(C_1-C_2)^2r}{2(2C_0+C_1+C_2)^2}
  \label{Utime}
\end{aligned}
\end{equation}

In addition to the typical $V^2$ term, the energy exhibits a time-dependent term that decays exponentially due to the time-varying potential. Taking the limit $r\rightarrow0$ reduces this equation to the metallic limit, while assuming that $\Delta q=0$ in Eq. (\ref{energydq}) yields the insulating limit. To compare with the experimental data and determine how the $\Delta f(V_{\rm bias},z)$ curves are modified, we substitute $t=((V(t)-V_{\rm 0})/\Delta V)T_s$ into Eq. (\ref{energydq}), obtaining 

\begin{equation}
\begin{aligned}
       U(V(t)) &= -\frac{1}{2}C_{\rm tot}(z) V_\textrm{bias}(t)^2\\
       &+\mathcal{A}(z)\exp\left({-\frac{V_\textrm{bias}(t)-V_{\rm 0}}{\Delta V}\frac{T_s}{\tau}}\right)+\mathcal{B}(z),
    \label{Uv}
\end{aligned}
\end{equation}

and 

\begin{equation}
\begin{aligned}
          \Delta f(V(t)) &=\frac{f_0}{2k_{\rm lever}}\left[ -\frac{1}{2}C_{\rm tot}^{\prime \prime}(z) V_\textrm{bias}(t)^2\right.\\
          &+\left.\mathcal{A}^{\prime\prime}(z)\exp\left({-\frac{V_\textrm{bias}(t)-V_{\rm 0}}{\Delta V}\frac{T_s}{\tau}}\right)+\mathcal{B}^{\prime\prime}(z)\right].
    \label{freqfitdiel}
\end{aligned}
\end{equation}
This expression shows that the $\Delta f(V_{\rm bias},z)$ relation is not purely parabolic anymore, but has an additional exponentially decaying term that depends on the voltage. If the conductivity is much larger than the typical measurement time ($\tau<<t_a$) the charge is able to reorganize 'instantaneously'. On the contrary, if the conductivity is much smaller ($\tau>>t_a$), charges do not move at all. For both cases the problem can be treated as static. However, this is not always the case, and when $\tau$ is comparable with the acquisition time the exponentially decaying term related to the charge dynamics will be non-negligible.

 As mentioned above, a small deviation from the parabolic behavior is already observed for GO (Fig. \ref{parabdiel}(b)), although its conductivity seems to be too low to appreciate the exponentially decaying contribution. Hence, to confirm its existence and magnify its effect, we slightly increase the material's conductivity by lightly reducing GO (GO100 sample). Moreover, as this effect becomes more prominent when the system is abruptly displaced from equilibrium, its influence should be most relevant at the beginning of each experimental $V_\textrm{bias}$ sweep. We recall that the voltage is swept only in the forward direction and that between each $V_\textrm{bias}$ sweep, the tip is kept at $V_\textrm{bias} = V_\textrm{KPFM}$, allowing the system to return to electrostatic equilibrium. 

\begin{figure}[h]
    \centering
    \includegraphics[width=8cm]{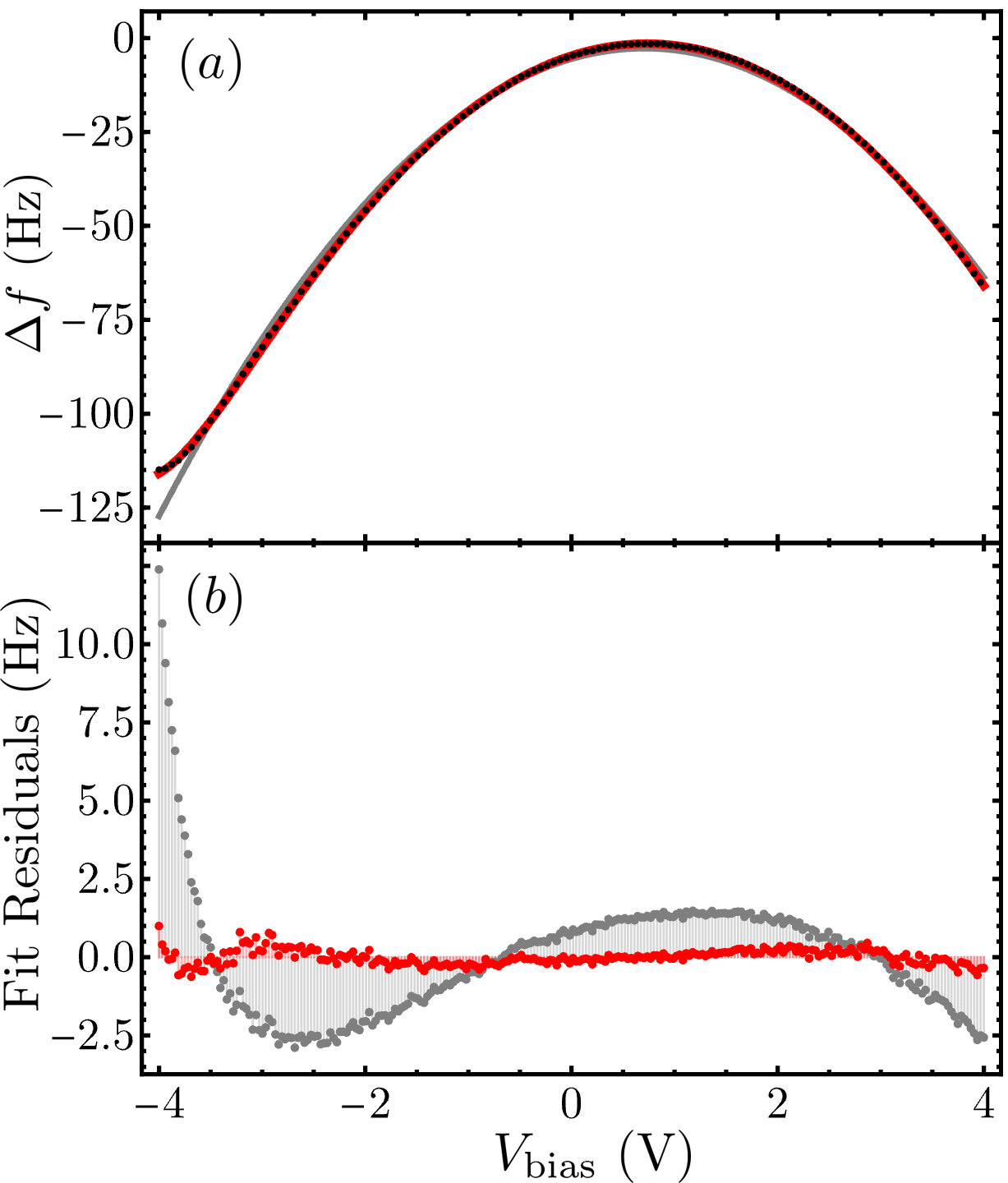}
    \caption{(a) $\Delta f (V_\textrm{bias},z)$ curve for GO100 at $z=15 $ nm. The solid gray line represents the fit to Eq. (\ref{freqfit}), while the solid red line represents the fit to Eq. (\ref{freqfitdiel}).  Every other point has been alternantly removed as otherwise the red curve is difficult to appreciate. (b) Residuals of the fits shown in (a). No points have been removed from this curve.}
    \label{parabGO100}
\end{figure}

A representative frequency shift curve for GO100 at RT, along with the parabolic fit (solid gray line) and the fit to Eq. (\ref{freqfitdiel}) (solid red line), is shown in Fig. \ref{parabGO100}(a) As evident from the data, and further confirmed by the fit residuals in Fig. \ref{parabGO100}(b), a parabolic model alone is insufficient to fit the data. In contrast, Eq. (\ref{freqfitdiel}) accurately reproduces the behavior, demonstrating that the exponentially decaying term is essential for capturing the details observed at the start of the voltage sweep.

\begin{figure}[h]
    \centering
    \includegraphics[width=8cm]{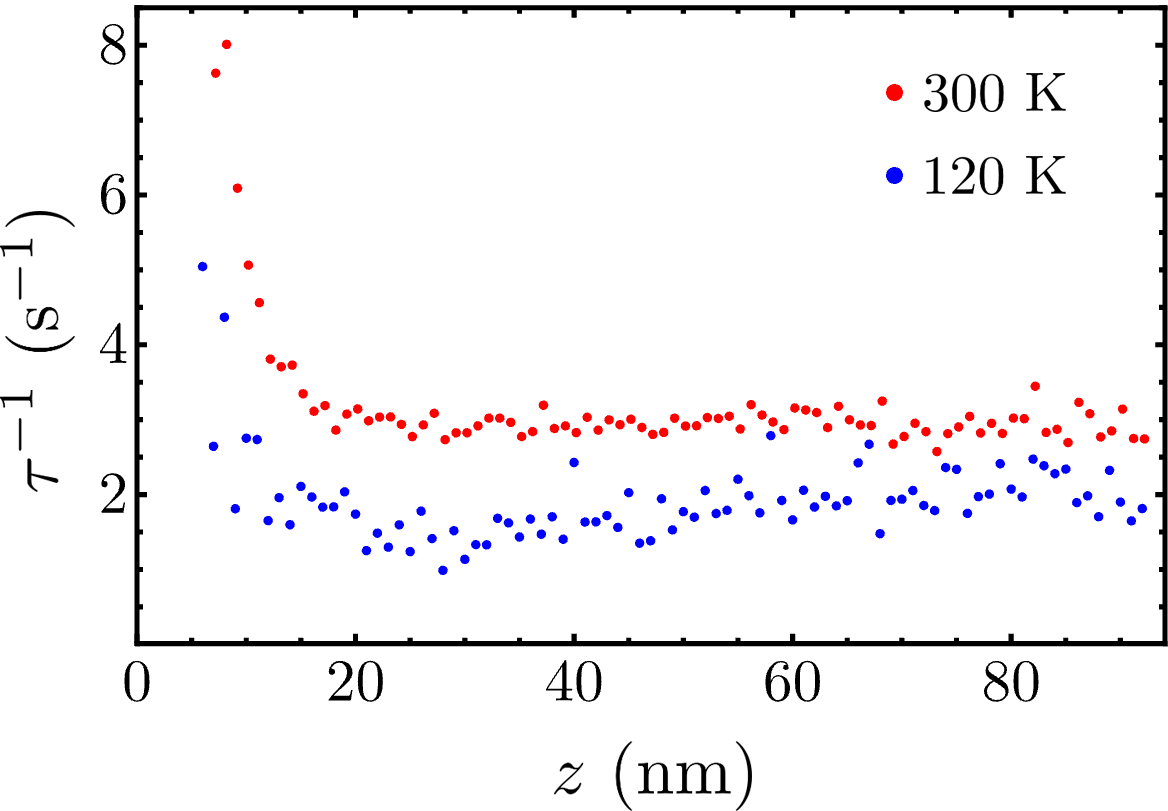}
    \caption{$\tau^{-1}$ vs. $z$ for sample GO100 at $300$ K and $120$ K.} 
    \label{TauGO100}
\end{figure}

By repeating the same experiments on GO100 at $120$ K, we expect a smaller value for $\tau^{-1}$ given that $\tau^{-1}\propto\sigma_s$ and conductivity increases with temperature in this material \cite{eda_insulator_2009, cheruku_variable_2018-2}. This is confirmed in Fig. \ref{TauGO100} where the values obtained for $\tau^{-1}$ at RT and $120$ K are shown. The conductivity of GO100 is estimated using Eq. (\ref{decay}) with $\varepsilon_1+\varepsilon_2\approx 5\varepsilon_0$ and considering $k \approx 1/R_\textrm{tip}$ obtaining $\sigma_s \sim  10^{-18}$ S, that is in good agreement with previously reported values \cite{eda_chemically_2010}. Finally we recall that to derive Eq. (\ref{freqfitdiel}), we have assumed large tip-sample distances and a $z$-independent $\tau$. Fig. \ref{TauGO100}, shows that this approximation is valid for $z \gtrapprox R_{tip}$, while it fails for smaller tip-sample distances. This dependence on $z$ is not unexpected as $\tau$ depends on the tip-sample distance through the capacitances $C_{1}$ and $C_{2}$. However, to reproduce its tendency we need to consider also the effect of multiple decay modes in our equations. Even more, at those small distances, the tip itself may influence the system's dynamics. This consideration significantly increases the complexity of the model and will be addressed in future works. 

\section{Conclusions}
In this study, we demonstrate that conductivity, particularly surface conductivity, may play a crucial role in the signals measured with AFM.   When surface conductivity is non-negligible, the charge dynamics induced by the presence of the tip must be carefully considered to ensure accurate interpretation of the data. This is especially significant when characterizing the electrical properties of 2D materials deposited on insulating substrates.

Our findings indicate that, in the high-conductivity limit, capacitance curves can be modeled as a semi-infinite metallic system, scaled by a parameter dependent on the material's conductivity. This approach offers an alternative method for indirect, contact-free conductivity measurements while also revealing potential size effects. In contrast, in the low-conductivity limit, the most striking phenomenon is the deviation from the expected parabolic dependence of the electrostatic potential on the applied voltage. This deviation arises because the characteristic charge relaxation time of charge carriers is comparable to or smaller than the typical measurement timescale.

Additionally, we emphasize that AFM electrostatic-related channels, beyond just the electrostatic potential, may be sensitive to surface conductivity. When an AC bias voltage, $V(t) = V_{\rm DC} + V_{\rm AC} \cos(\omega_{\rm elec} t)$, is applied, the metallic or dielectric response depends not only on the intrinsic material properties but also on the product $f_{\rm elec} \tau$. In particular, when $f_{\rm elec} \tau \approx 1$, surface potential measurements may not correspond to an equilibrium state. Moreover, the $2\omega_{\rm elec}$ channel, commonly associated with the system's capacitance and used to determine the relative permittivity of dielectric materials, proves to be an effective tool for distinguishing between different types of flakes in GO/RGO mixtures \cite{navarro-rodriguez_exploring_2024}. Our model suggests that this contrast primarily arises from the influence of surface conductivity.

\begin{acknowledgments}
The authors thank the funding agencies MCIN/AEI/10.13039/501100011033 through projects PID2022-139191OB-C31 and PDC2023-145906-100 "A way of making Europe”; M.N work was financed by the grant PID2019-104272RB-C52/PRE2020-094503 funded by MCIN/AEI/ 10.13039/501100011033 and by “ESF Investing in your future”.
\end{acknowledgments}
\newpage
\setcounter{equation}{0}
\setcounter{section}{0}
\setcounter{figure}{0}
\setcounter{table}{0}
\setcounter{page}{1}
\makeatletter
\setlength{\parindent}{0pt}
\renewcommand{\thefigure}{S\arabic{figure}}
\renewcommand{\thetable}{S\Roman{table}}
\renewcommand{\thesection}{S\Roman{section}}
\include{SMaterial}
\bibliography{bibliographymain}
\end{document}

%% file: SMaterial.tex
\title{Supplemental material to the manuscript\\
Modelling of time-dependent electrostatic effects and AFM-based surface conductivity characterization}
\maketitle
\section{Resonance frequency and $k_\textrm{lever}$ as a function of T}
\begin{figure}[h]
    \centering
    \includegraphics[width=10cm]{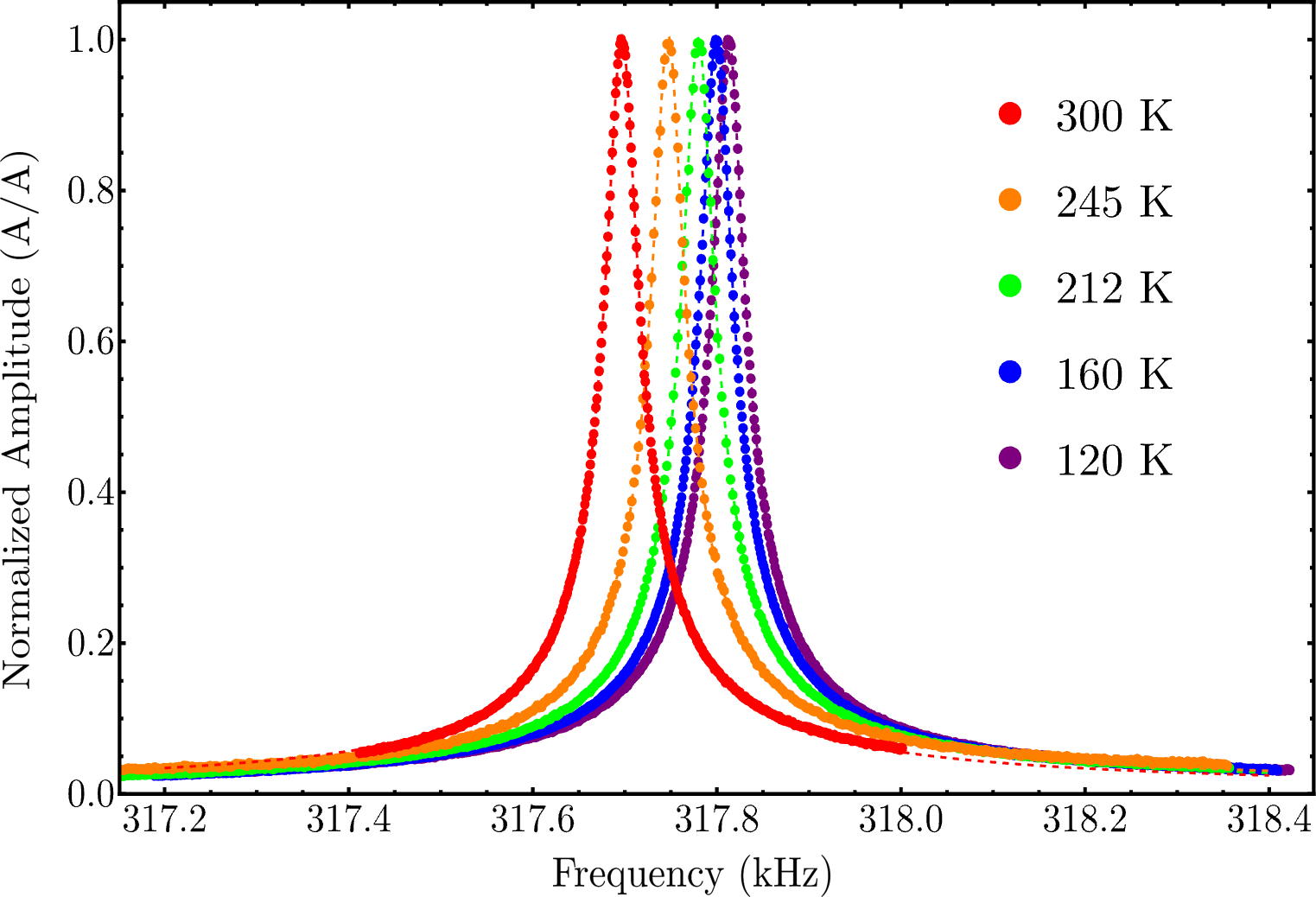}
    \caption{Resonance curve of the cantilever as a function of the temperature}
    \label{resonancefreq}
\end{figure}

The cantilever's resonance frequency is obtained from the fit to a Lorentzian curve at each temperature. From the resonance frequency, $f_0$, the force constant is calculated as 

\begin{equation}
    k_\textrm{lever}=0.25\rho_\textrm{Si}V_\textrm{cantilever}(2\pi f_0)^2,
\end{equation}
according to \cite{garcia_amplitude_2010,gysin_temperature_2004}. Where $\rho_\textrm{Si}$ is the density of silicon and $V_\textrm{cantilever}$ is the cantilever volume. The factor of $0.25$ comes from the point mass approximation. The volume is calculated from the manufacturer's specifications (length $125$ $\mu$m, width $30$ $\mu$m and thickness $4$ $\mu$m). 

\begin{table}[h]
\caption{\label{tab:table1}%
Resonance frequency and force constant of the cantilever at each temperature.
}
\begin{tabular}{ccc}
\hline \hline
Temperature (K) & $f_0 $ (kHz) & $k_\textrm{lever}$ (nN/nm) \\
\hline
$300$        &  317.697   & $34.800$                   \\
$245$        &  317.747   & $34.812$                   \\
$212$        &  317.780   & $34.819$                   \\
$160$        &  317.799   & $34.823$                   \\
$120$        &  317.813   & $34.826$                  \\
\hline \hline
\end{tabular}
\end{table}
\newpage

\section{$C^{\prime\prime}(z)$ for RGO at different amplitudes}

\begin{figure}[h]
    \centering
    \includegraphics[width=10cm]{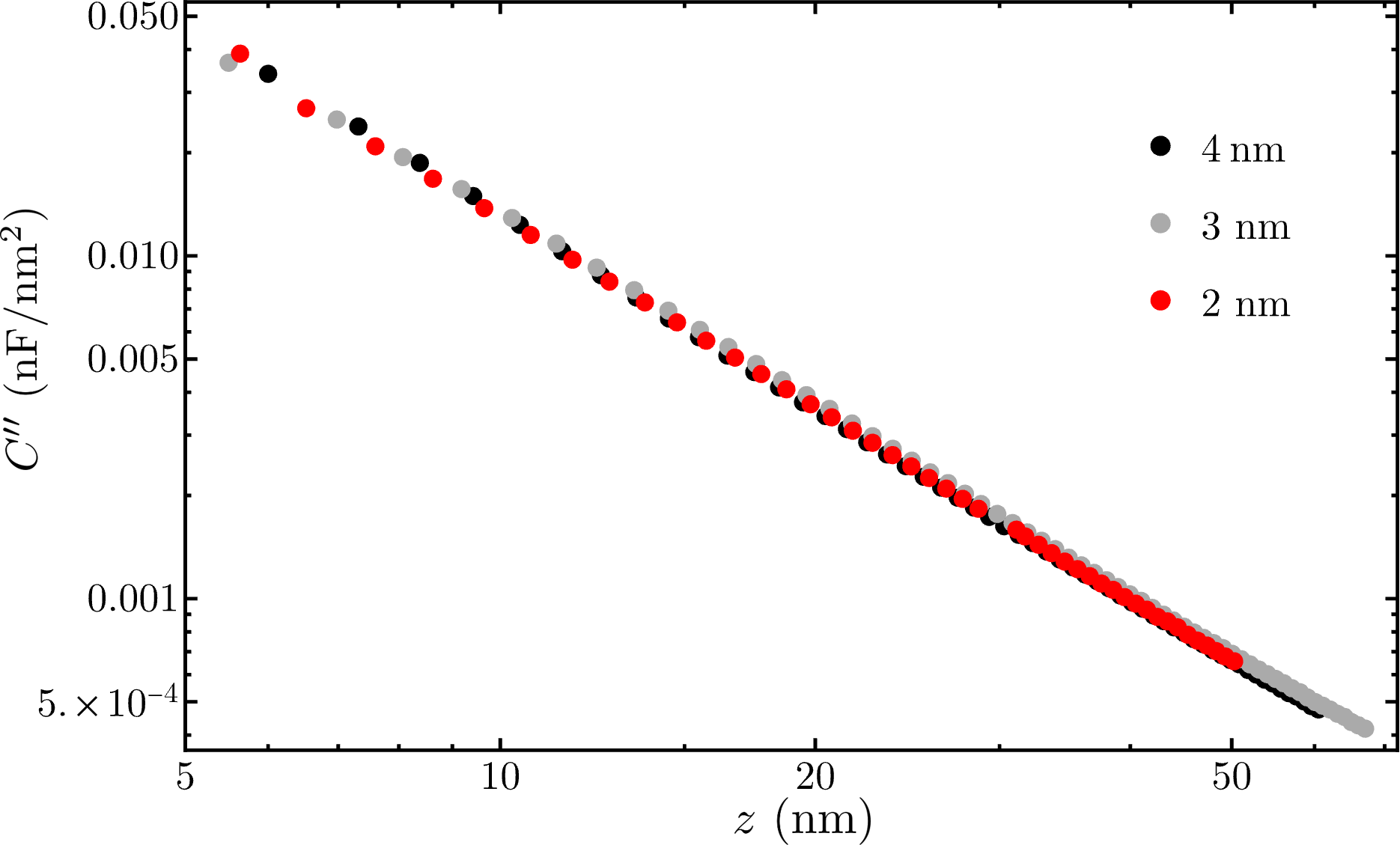}
    \caption{$C^{\prime \prime}(z)$ for three different oscillation amplitudes at room temperature measured on RGO. The curves match almost exactly, so this amplitude range does not have an influence in the measurements.}
    \label{rGOAmp}
\end{figure}

\section{Simulation of $C^{\prime\prime}(z)$ curve of a thin film on top of an insulating substrate}

\begin{figure}[h]
    \centering
        \includegraphics[width=12cm]{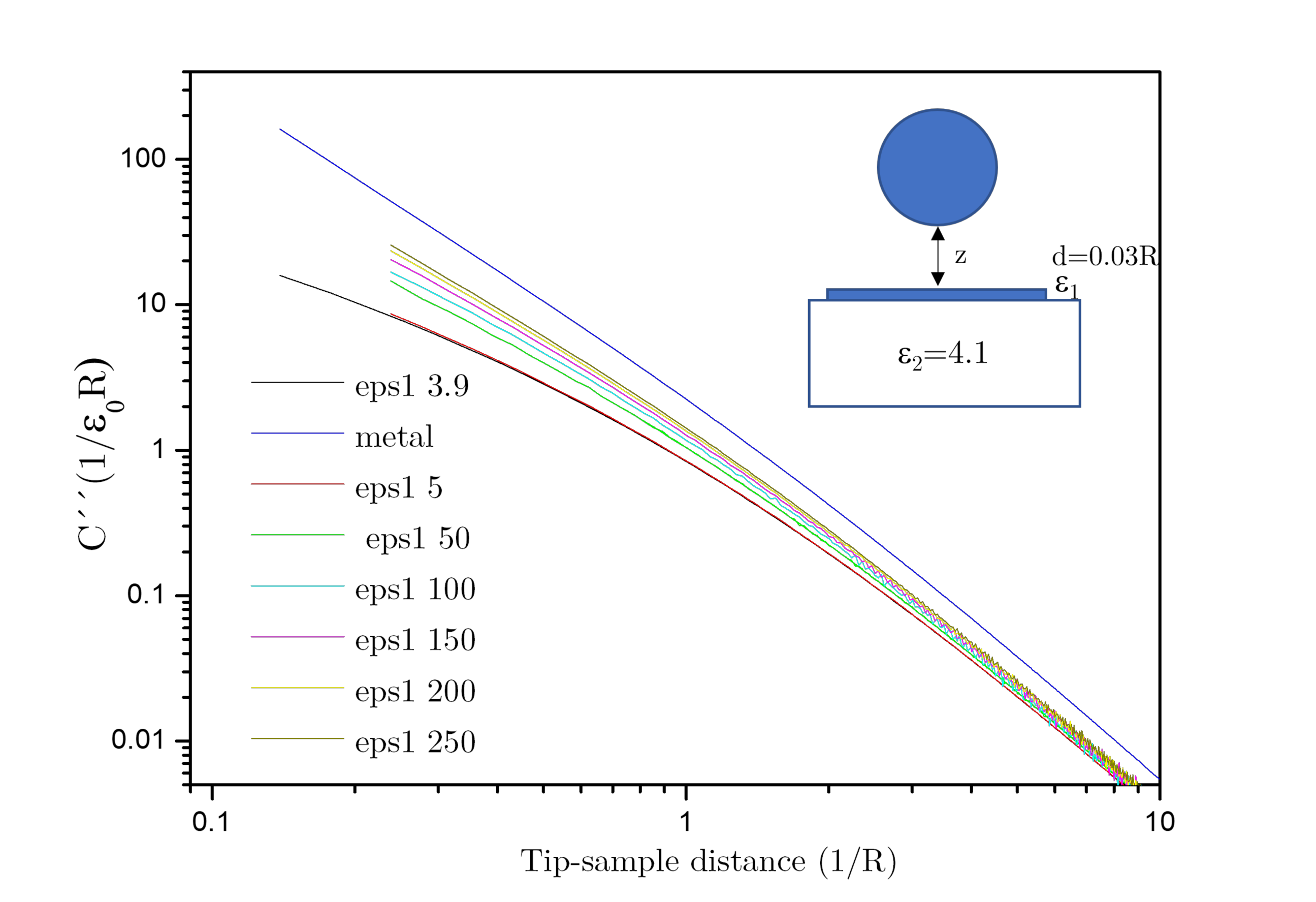}
    \caption{Simulated $C^{\prime\prime}(z)$ curves for an spherical tip placed on top of a thin film of thickness $1$ nm and relative permittivity $\epsilon_1$ deposited on a semi-infinite dielectric medium of $\epsilon_{2}=4.1$. Different curves represent various values of $\epsilon_1$. The metallic limit  ($\epsilon_1=\infty$ shown as a blue line) has been included to is included to highlight that, even for large values of $\epsilon_1$, the response remains significantly far from the metallic case due to the small film thickness.}
    \label{cpp_epsr}
\end{figure}

\section{Voltage independent term for rGO as a function of the temperature}

\begin{figure}[h]
    \centering
    \includegraphics[width=10cm]{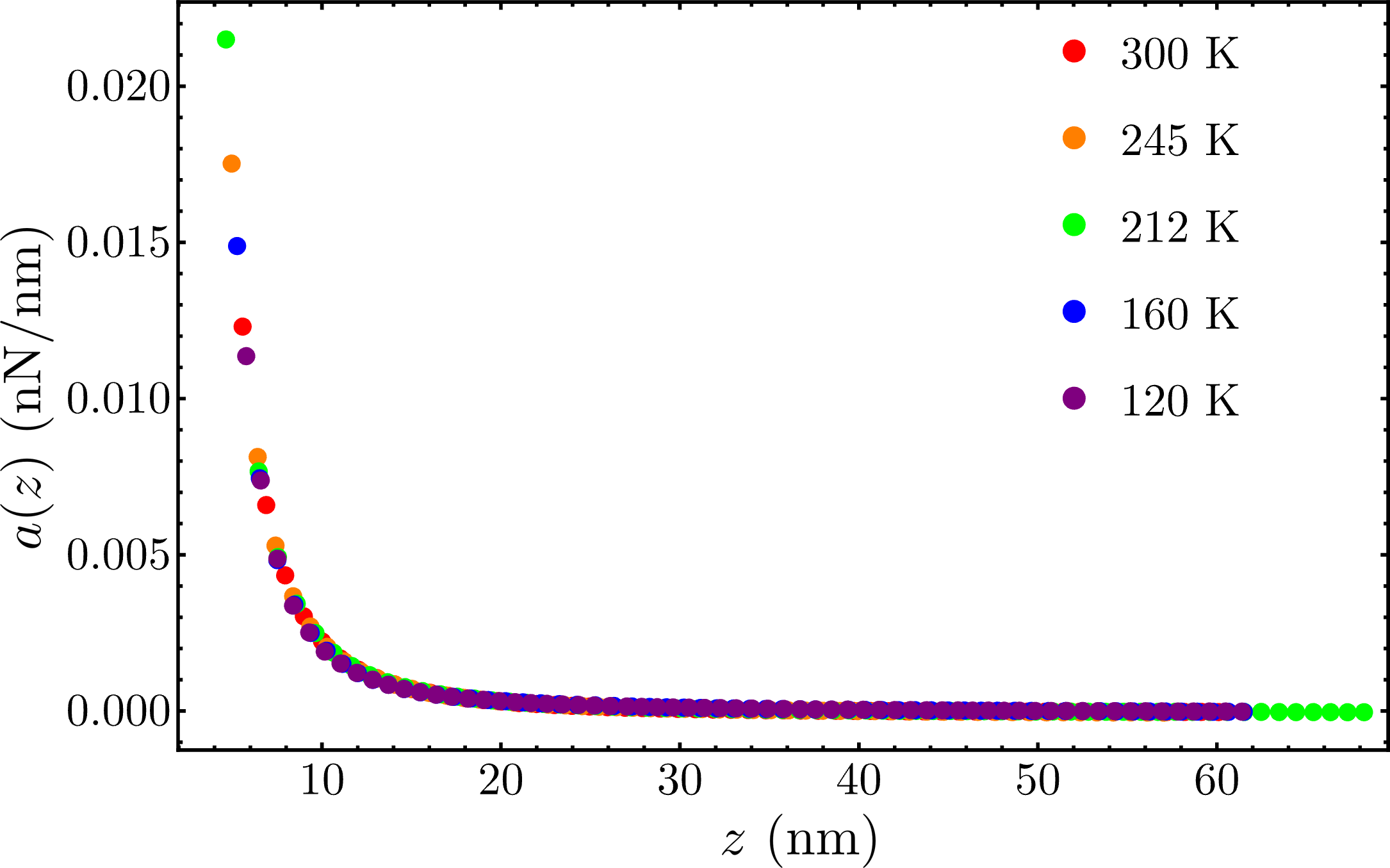}
    \caption{Comparison of the voltage independent term obtained from the fits as a function of the temperature for RGO.}
    \label{vindeprGO}
\end{figure}

The voltage independent term $a(z)$ obtained from the fits to a parabola for different temperatures are shown in Fig. \ref{vindeprGO}. 

\section{$C^{\prime\prime}(z)$ fit to a metallic model for RGO at different temperatures}
\begin{figure}[h]
    \centering
    \includegraphics[width=10cm]{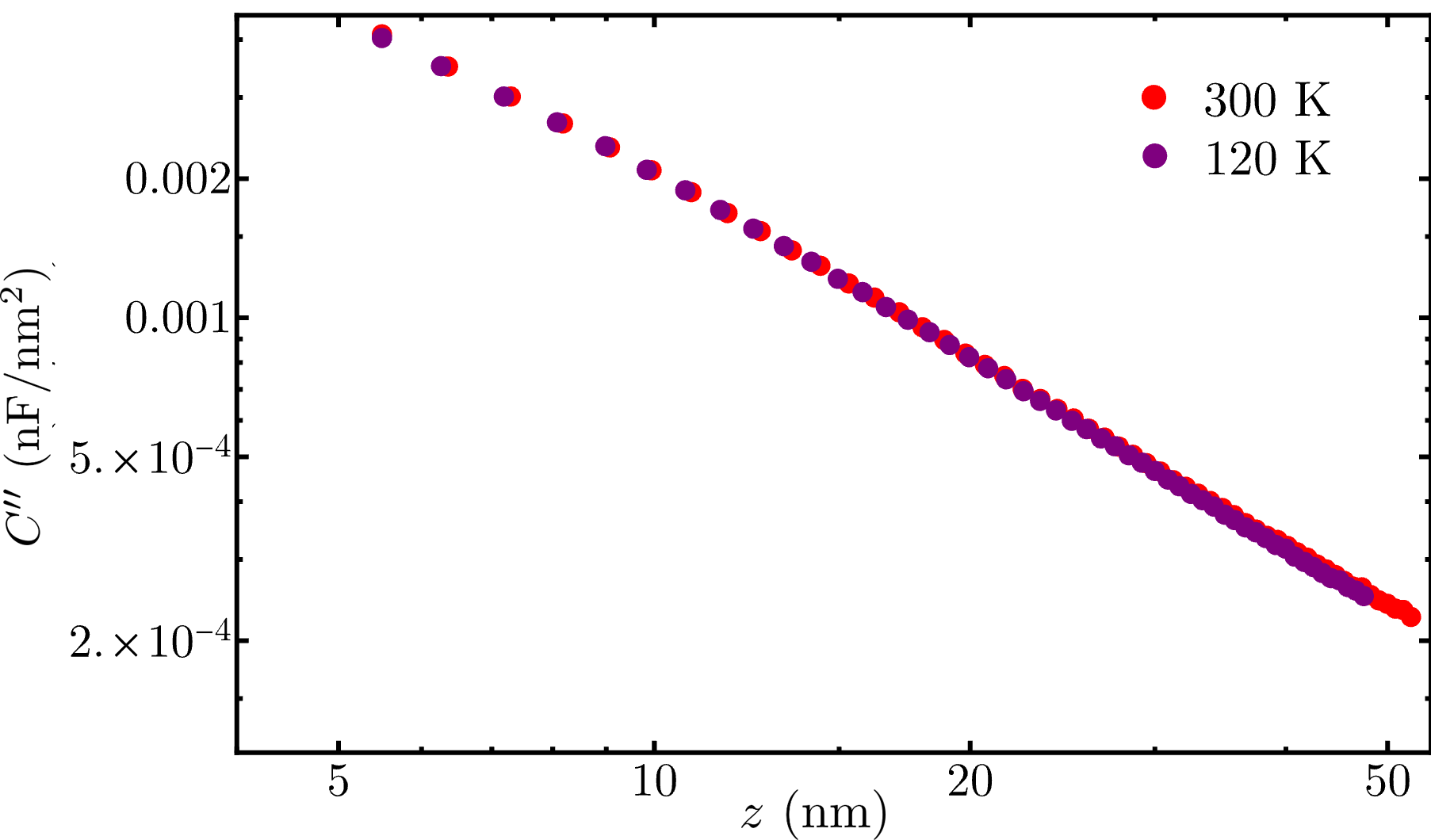}
    \caption{Comparison of the term $C^{\prime\prime}(z)$ at RT and 120K on the SiO$_2$ substrate performed at the beginning (red) and at the end of the experiments (purple).}
    \label{SiO2_T}
\end{figure}

For the fitting procedure, we analyzed two different experimental sets, with oscillation amplitudes of $3$ and $4$ nm. These sets were fitted simultaneously (only the $A=4$ nm set is shown in the main text for clarity), with the parameters to be determined being $\alpha$, the tip radius $R_{\rm tip}$, the initial tip-sample distance $z_0$, and the cone weight $b$. All curves were assumed to have the same tip radius. The integrity of the tip radius was verified by confirming that the curvature measured on the SiO$_2$ substrate remained unchanged at the start and end of the experiments (Fig. \ref{SiO2_T}). For a given temperature, curves with different oscillation amplitudes were constrained to share the same $\alpha$ value. Each curve had a single independent degree of freedom: the initial tip-sample distance ($z_0$), which accounts for slight variations in the initial tip-sample distance. However, all values remained within $z_0=6\pm1.5$ nm. 

\newpage

\section{$C^{\prime\prime}(z)$ for SiO$_2$ at RT and 120K}

\section{Estimation of $\varepsilon_{\textrm{SiO}_2}$}
The dielectric permittivity of the SiO$_2$ substrate ($\varepsilon_{\textrm{SiO}_2}$) has been estimated by fitting the corresponding $C^{\prime\prime}(z)$ experimental curves to the model used in \cite{orihuela_localized_2016}. Briefly, the apex contribution is calculated numerically considering a infinite dielectric while the cone contribution is included considering a thin film \cite{labardi_extended_2015}.  

\begin{figure}[h]
    \centering
    \includegraphics[width=8cm]{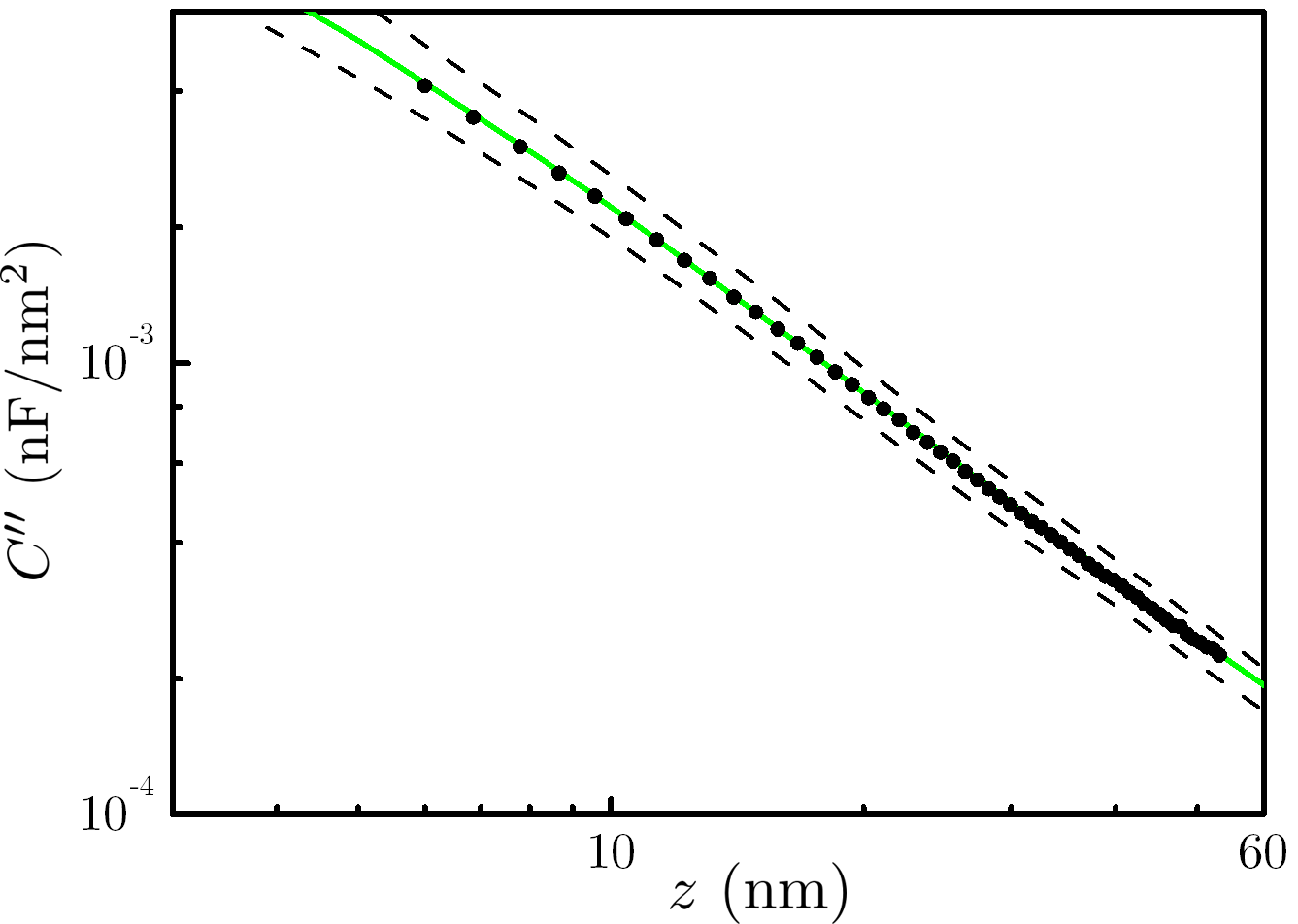}
    \caption{Fit of the term $C^{\prime\prime}(z)$ for the SiO$_2$. The parameters used were $R_\textrm{tip}=28$ nm and a SiO$2$ thickness of $300$ nm. The best fit yields $\varepsilon_{\rm{SiO}_2}=4.1\varepsilon_0$ The dashed lines represent the curves for $\varepsilon_r=3.5$ and $\varepsilon_r=4.5$, serving as a confidence interval.} 
    \label{SiO2fit}
\end{figure}